\documentclass[default]{aastex701}
\usepackage{graphicx} 
\usepackage{dblfloatfix}      
\usepackage{placeins}         
\usepackage{gensymb}
\usepackage[table]{xcolor}
\usepackage{xcolor}
\usepackage{soul}
\usepackage{amsmath}
\usepackage{CJK}
\usepackage{hyperref}
\usepackage{tikz}
\usepackage{multirow}
\usepackage{soul}
\usepackage[percent]{overpic}

\newcommand{\rr}{\boldsymbol r}
\newcommand{\p}{\boldsymbol p}
\newcommand{\Ks}{\boldsymbol{K}_s}
\newcommand{\vsw}{\boldsymbol{v}_{\rm sw}}
\newcommand{\B}{\boldsymbol{B}}
\newcommand{\vv}{\boldsymbol{v}}
\newcommand{\e}{\boldsymbol{e}}

\begin{document}

\title{On modelling the 2017 galactic cosmic ray depression}

\author[0000-0001-9521-3874]{O.P.M. Aslam}\email[]{muhammedaslam.ottupara@glasgow.ac.uk}
\affiliation{School of Mathematics and Statistics, University of Glasgow, Glasgow G12 8QQ, United Kingdom}

\author[0000-0003-2297-9312]{D. MacTaggart}\email[show]{david.mactaggart@glasgow.ac.uk}
\affiliation{School of Mathematics and Statistics,
University of Glasgow,  Glasgow G12 8QQ, United Kingdom}
\affiliation{Department of Physics, University of Trento, Povo, Italy}

\author[0000-0003-0793-7333]{M.S. Potgieter}\email[]{t@t}
\affiliation{Institute for Experimental and Applied Physics (IEAP), Christian-Albrechts-University of Kiel, Kiel, Germany}



\author[0000-0001-5844-3419]{M.D. Ngobeni}\email[]{t@t}
\affiliation{Department of Physical and Earth Sciences, Sol Plaatje University, Kimberley, South Africa}

\author[0000-0002-0605-1306]{I.I. Ramokgaba}\email[]{t@t}
\affiliation{School of Physical and Chemical Sciences, North-West University, Mmabatho, South Africa}
\affiliation{Centre for Space Research, North-West University, Potchefstroom, South Africa}

\begin{abstract}
In the second half of 2017, as Solar Cycle 24 approached its minimum, a pronounced and rigidity‑dependent depression in the galactic cosmic ray (GCR) proton flux was observed using AMS‑02. This event was driven by a sequence of strong coronal mass ejections (CMEs). In this work, we model the modulation of GCR protons during this period using a steady‑state Parker transport framework with parameters constrained by AMS‑02 observations. Treating the event as a purely long‑term modulation interval yields excellent agreement with the observed proton intensities, but implies an inferred rigidity dependence of particle mean free paths that is inconsistent with basic turbulence theory, corresponding to an effective inversion of the scattering regime. We argue that this behaviour arises from the ability of highly flexible transport parameterizations to reproduce integrated modulation signatures through correlated adjustments of diffusion coefficients during transient events. To address this inconsistency, we relax the commonly adopted assumption of a time‑independent ratio of perpendicular to parallel diffusion coefficients and introduce a physically motivated reduction of this ratio during the CME‑dominated interval. This modification preserves the quality of the spectral fits whilst restoring the expected rigidity ordering of mean free paths. Our results indicate that allowing the perpendicular‑to‑parallel diffusion ratio to vary in time provides an effective means of constraining transport solutions during the 2017 depression and may help to resolve similar inconsistencies reported in other GCR modulation studies.
\end{abstract}


\section{Introduction}

Galactic cosmic rays (GCRs) that enter the heliosphere are modulated by solar activity \citep{2013LRSP...10....3P}. The clearest signature of this modulation is that of the solar cycle, in which GCR flux behaves in an anti-phase manner with respect to the solar cycle, i.e. the GCR flux peaks at solar minimum and troughs at solar maximum. In the second half of 2017, more specifically from 23 June to 03 December 2017 (Carrington rotations (CR) 2192 - CR2198), the Alpha Magnetic Spectrometer ({AMS-02}) \citep{2021PhR...894....1A}, located on the International Space Station, revealed a large and trend-reversing  depression in the GCR proton ($p$) flux for a range of different rigidity bins \citep{2021PhRvL.127A1102A}.  This was a period when Solar Cycle 24 was approaching its minimum. The expected conditions were disturbed, however, by the presence of a highly active magnetic source on the Sun, producing a series of strong flares and coronal mass ejections (CMEs), that persisted for five solar rotations, with Active Region (AR) labels: AR12665 $\rightarrow$ AR12670 $\rightarrow$ AR12673 $\rightarrow$ AR12682 $\rightarrow$ AR12685 \citep[for more details, see][]{2025ApJ...981..174A}. The production of strong flares and CMEs as the cycle approaches a minimum is a feature that has also been observed for other solar cycles \citep[e.g.][]{Tan2025}. From the perspective of understanding GCR modulation, this activity poses interesting questions about how to model GCR transport. In particular, what are the consequences of an interval of high solar activity during what are otherwise conditions close to those of solar minimum? 

Although not focussing on the  2017 GCR depression \emph{per se}, several works have studied GCR modulation across longer periods that include the depression. \citet{2023ApJ...953..101A} reconstruct the modulation of GCR protons and antiprotons from 2006 to 2019. To study long-term modulation effects, they make use of a well-tested steady-state model \citep{2017AdSpR..60..848P} which captures modulation behaviour on a minimum time scale of one Carrington (or Bartels) rotation and is suitable for long-term modulation studies. The model results of \citet{2023ApJ...953..101A} clearly show the 2017 GCR depression for both protons and antiprotons. The parameterization of the model is constructed based on fitting data from AMS-02 and PAMELA \citep{2018PhRvL.121e1101A,2021PhRvL.127A1102A, 2013ApJ...765...91A, 2018ApJ...854L...2M}. \citet{CortiICRC2023}, employing a similar steady-state transport model combined with a neural‑network‑accelerated parameter exploration algorithm, reports results that highlight potential physical inconsistencies in the inferred rigidity dependence of transport coefficients during the declining phase of the cycle. In particular, their work suggests that model parameterizations capable of reproducing observed intensities may imply an inverted effective rigidity dependence of scattering, in apparent tension with standard expectations from basic turbulence theory. Time‑dependent modelling efforts have revealed similar behaviour \citep[e.g.][during the solar minimum of Cycle 24]{2021ApJS..257...48S}, indicating that this issue is not solely a consequence of steady‑state assumptions.

The purpose of this work is not to introduce a new cosmic ray modulation model, but to examine the physical consistency of standard long‑term modulation parameterizations when applied to mid-term, CME‑dominated intervals, thus providing a physically consistent description of the modulation of GCR protons during the 2017 depression. We show that treating this event as a purely long‑term modulation episode yields good agreement with AMS‑02 intensities, but implies an inferred rigidity dependence of particle mean free paths that is difficult to reconcile with basic expectations from turbulence theory, resulting in similar issues found by \citet{CortiICRC2023} and \citet{2021ApJS..257...48S}. We argue that this behaviour reflects the nonlinearity of the model, rather than evidence for unusual or inverted heliospheric turbulence. In particular, the nonlinear model supports different parameterizations that can match the observational data. 

Motivated by the hybrid nature of the 2017 depression (as a period of intense solar activity between near-solar minimum conditions), we relax the commonly adopted assumption of a time‑independent ratio of perpendicular to parallel diffusion coefficients. By allowing this ratio to decrease during the CME‑dominated depression, we obtain transport solutions that preserve agreement with AMS‑02 observations while restoring the physically consistent rigidity ordering of the mean free paths. This modification acts as a physically motivated constraint on the solution space, limiting compensating parameter adjustments that can otherwise mask inconsistent transport behaviour in events like the 2017 depression. The results suggest that introducing time variability in the perpendicular‑to‑parallel diffusion ratio may provide a useful diagnostic and corrective approach for interpreting similar modulation intervals reported in recent studies.

The paper is structured as follows. In Section 2 we summarize the elements of the transport model and the parameterizations employed. Section 3 presents results obtained by modelling the 2017 depression using a conventional long‑term modulation approach and highlights the resulting physical inconsistency. In Section 4, we introduce the hybrid‑interval parameterization and demonstrate how it restores physically consistent transport behaviour. Section 5 summarizes the main conclusions and discusses implications for future GCR modulation studies.

\section{Elements of modelling GCR modulation}

The four major modulation mechanisms are: outward convection by the solar wind (SW), adiabatic deceleration due to the expanding SW (adiabatic cooling), diffusion along and across the heliospheric magnetic field (HMF) and drifts related to the gradient and curvature of the HMF and the heliospheric current sheet (HCS). These modulation processes are responsible for altering the differential intensities of cosmic rays as a function of energy, position and time, and were first combined by \citet{1965P&SS...13....9P} into a comprehensive transport equation \citep[see also][]{1967ApJ...149L.115G}. The basic transport equation follows from the equation of charged particles motion in large- and small-scale fluctuating magnetic fields, and averages over the pitch and phase angles of particle propagation, with the assumption that GCRs are approximately isotropic.  Within a heliocentric spherical coordinate system that rotates with the Sun, the time-dependent transport equation, incorporating explicit drift \citep{1977ApJ...213..861J}, is given by

\begin{equation}
\underbrace{\frac{\partial f}{\partial t}}_{\rm (a)} = \underbrace{- \vsw \cdot \nabla {f}}_{\rm (b)} - \underbrace{\langle \vv_D \rangle \cdot \nabla {f}}_{\rm (c)} + \underbrace{\nabla \cdot (\Ks \cdot \nabla \it{f})}_{\rm (d)} + \underbrace{\frac {1}{3} (\nabla \cdot \vsw) \frac {\partial f} {\partial  \ln  p}}_{\rm (e)} + \underbrace{Q}_{\rm (f)} \label{Eq-TPE}
\end{equation}
where $f (\rr, p, t)$ is the omnidirectional GCR distribution function, where $p$ is the magnitude of the particle momentum, $t$ is time and $\rr$ is the position vector in 3D with coordinates $(r, \theta, \phi)$ specified in a heliocentric spherical coordinate system where the equatorial plane is at a polar angle of $\theta = 90^{\circ}$. 


In equation (\ref{Eq-TPE}), the individual terms are: (a) time-dependent changes in the GCR distribution function; (b) outward convection caused by the radially expanding SW, with velocity $\vsw$, within the co-rotating frame; (c) GCR gradient and curvature drift motions in the global HMF in terms of the averaged pitch angle guiding center drift velocity $\langle \vv_D \rangle$; (d) spatial diffusion caused by the irregular HMF through the symmetric diffusion tensor $\Ks$; (e) adiabatic energy changes (deceleration or acceleration) determined by the SW divergence $\nabla \cdot \vsw$ (if $\nabla \cdot \vsw>0 $, adiabatic energy loss (cooling) occurs as is the case in most of the heliosphere, except inside the heliosheath where we assume that $\nabla \cdot \vsw = 0$; see also \citet{2006ApJ...640.1119L}) and (f) possible additional sources of cosmic rays within the heliosphere (e.g. Jovian electrons). The heliosheath in the model is the region between the heliopause (HP), assumed to be at 122 AU, and the SW termination shock (TS), which moves around with solar activity \citep[see also][]{2015ApJ...799..223M}. 
For more details about heliospheric transport processes and GCR modulation in general, there are several review papers, including \citet{1998SSRv...83..147P}, \citet{1999AdSpR..23..415F}, \citet{2013LRSP...10....3P,2013SSRv..176..165P,2017AdSpR..60..848P} and \citet{2022SSRv..218...42R}. In this study, we are interested in the behaviour of GCRs over a period spanning several solar rotations, for which we take our minimum time scale to be a Carrington rotation. We thus make the simplifications $\partial f/\partial t = Q = 0$. Doing this means that we seek the dominant balance in equation (\ref{Eq-TPE}) averaged over a Carrington rotation. In other words, the outward convection of the SW (b) must be balanced by a combination of drift, diffusion and adiabatic cooling (c-e) for each Carrington rotation. This approach is consistent with previous modulation studies that address mid-to-long‑term variability using rotation‑averaged parameters  \citep{2023ApJ...953..101A,CortiICRC2023}.

\subsection{\bf Diffusion}

The convection process in equation (1) is well understood and is based on the SW velocity, as described in detail by \citet[][]{2014SoPh..289..391P}.
As a result of turbulent irregularities and fluctuations in the HMF, GCRs undergo diffusion in the heliosphere through a process called pitch angle scattering, which can be described by {quasi-linear theory} (QLT) as introduced by \citet{1966ApJ...146..480J}. Such turbulence is generally interpreted either as waves \citep{1988JGR....93.2725S} or dynamical turbulence \citep[e.g.][]{1991ICRC....3..248B}, and is described by diffusion coefficients (DCs) that make up the elements of the tensor ${\Ks}=(\kappa_{ij})$, $1\le i,j\le 3$. These DCs can be determined from turbulence theory \citep[see e.g.][]{2020ApJ...894..121M} or be based on the compatibility studies between numerical modulation models and cosmic ray observations in the heliosphere, as reviewed by \citet{2017AdSpR..60..848P}.

Each DC is related to a mean free path (MFP) $\lambda$ by

 \begin{equation}\label{lambda}
 \kappa = \frac{v}{3} \lambda ,   
 \end{equation}
where  $v$ is the particle speed. For the case when $\kappa$ is the drift coefficient (see below for details), $\lambda$ is referred to as the drift scale ($\lambda_{D}$). 

In the context of our steady‑state approach, the DCs, MFPs, etc. should be interpreted as effective, Carrington‑averaged transport parameters rather than instantaneous or local turbulence quantities. They, therefore, encapsulate the cumulative influence of transient disturbances, such as CMEs, on GCR transport over the averaging interval, rather than directly representing microscopic coefficients.

\subsubsection{Parallel Diffusion}



To model the parallel (to the magnetic field) DC, we follow previous work \citep[e.g.][]{2014SoPh..289..391P} 
in adopting a rigidity-dependent expression that has been found to be a good approximation to QLT. For its spatial dependence, the parallel DC, $\kappa_{\parallel}$, is considered to be inversely proportional to the HMF magnitude. The general expression used in our study to describe it is given by

\begin{equation}\label{EqParallelDiffusion}
\kappa_{\parallel} = \kappa_{\parallel 0} \beta \frac {B_{0}}{B_m} \Bigg(\frac {P}{P_{0}}\Bigg)^{c_{1}} \left[ \frac {\Bigg(\frac {P}{P_{0}}\Bigg)^{c_{3}} + \Bigg(\frac {P_{k}}{P_{0}}\Bigg)^{c_{3}}}{ 1+ \Bigg(\frac {P_{k}}{P_{0}}\Bigg)^{c_{3}}} \right]^{\frac {c_{2 \parallel} - c_{1}}{c_{3}}}
\end{equation}
where $P$ is the rigidity, $B_m$ is the magnitude of the modified HMF (see below for further details), $\beta=v/c$ ($c$ is the speed of light) 
and $\kappa_{\parallel 0}$ is a scaling constant in units of $10^{22}$ cm$^{2}$s$^{-1}$, with the rest of the equation written to be dimensionless, with $P_{0} = 1.0$ GV, and $B_{0}$ = 1.0 nT, in order to preserve the units in cm$^{2}$ s$^{-1}$. Furthermore, $c_{1}$, $c_{2\parallel}$ and $c_{3}$ are dimensionless; $c_{1}$ is a power index that may change with time; $c_{2 \parallel}$ and $c_{1}$ determine the slope of the rigidity dependence, respectively, above and below a rigidity with the value $P_{k}$ which may also change with time and $c_{3}$ determines the smoothness of the transition. 
The rigidity dependence of $\kappa_{\parallel}$ is thus a combination of two power laws, $P_{k}$ determines the rigidity where the transition in the power laws occur and the value of $c_{1}$ determines the slope of the power law at rigidities below $P_{k}$.

\subsubsection{Perpendicular Diffusion  \label{perpendicular}}

The scattering of GCRs perpendicular to the HMF can be caused either as a result of the particles’ gyrocentres being displaced transversely relative to the mean HMF through scattering, or due to the random walk of the magnetic field lines themselves. These processes are collectively taken into account in numerical models via the perpendicular DC, $\kappa_{\perp}$. 
The coefficient $\kappa_{\perp}$ may be subdivided into two possibly independent coefficients describing perpendicular diffusion in the radial, $\kappa_{\perp r}$, and polar, $\kappa_{\perp \theta}$, directions. It has also been established that $\kappa_{\perp}$ plays a significant role in the modulation of GCRs \citep[e.g.][]{2000JGR...10518295P, 2000JGR...10518305F}. 

In the presence of parallel diffusion, a pure field line random walk scenario gives an insufficient description of perpendicular diffusion because particles sometimes retrace their paths after they back-scatter - a process that was not taken into account until the nonlinear guiding center (NLGC) theory of particle diffusion proposed by \citet{2003ApJ...590L..53M}. According to this theory, the process of perpendicular diffusion is a combination of a field line random walk, back-scattering from parallel diffusion and the transfer of particles across field lines due to the perpendicular complexity of the magnetic field \citep[see also][]{2004GeoRL..3110805B}. This theory for perpendicular diffusion was later improved by \citet{2006A&A...453L..43S}, who proposed an extended nonlinear guiding center theory (ENLGC), followed by a unified NLGC theory (UNLGC) in \citet{2010ApJ...720L.127S}. For overviews of nonlinear GCR diffusion theories and comparisons with observations, the reader is directed to \citet{2009ASSL..362.....S}, \citet{2010JGRA..115.3103P} and \citet{2022SSRv..218...33E}. 

According to the NLGC approach, perpendicular MFPs are calculated using parallel MFPs. 
In our work, for rigidities above $P_{k}$, $\kappa_{\perp}$ is scaled with a slightly weaker dependence compared to $\kappa_{\parallel}$, which, after converting to MFPs, results in a non-constant $\lambda_{\perp}/\lambda_{\parallel}$ ratio \citep[e.g.][]{2004ApJ...615..805S, 2010JGRA..115.3103P}. This translates to using different values for $c_{2}$ in equation (\ref{EqParallelDiffusion}), namely $c_{2\parallel}$ for $\kappa_{\parallel}$ and $c_{2\perp}$ for $\kappa_{\perp}$. 

Observations from the Ulysses spacecraft \citep[][and references therein]{2006SSRv..127..117H} also revealed that the latitude dependence of GCR proton fluxes is significantly weaker than predicted by classical drift models \citep[][]{1993ICRC....3..457P}, which led \citet{1995ICRC....4..680K} to propose the concept of an anisotropic $\kappa_{\perp}$, where $\kappa_{\perp \theta}$ $>$ $\kappa_{\perp r}$ in the off-equatorial regions \citep[see also][]{2000JGR...10518295P,2000JGR...10527447B}. 

The perpendicular DCs are given by
\begin{equation}
 \kappa_{\perp r} = \kappa^{0}_{\perp r} \kappa_{\parallel 0} \beta \frac {B_{0}}{B_m} \Bigg(\frac {P}{P_{0}}\Bigg)^{c_{1}} \left[ \frac {\Bigg(\frac {P}{P_{0}}\Bigg)^{c_{3}} + \Bigg(\frac {P_{k}}{P_{0}}\Bigg)^{c_{3}}}{ 1+ \Bigg(\frac {P_{k}}{P_{0}}\Bigg)^{c_{3}}} \right]^{\frac {c_{2 \perp r} - c_{1}}{c_{3}}},
\label{Eq-perpDiffR}
\end{equation}
and
\begin{equation}
\kappa_{\perp \theta} = f_{\perp \theta}\kappa^{0}_{\perp \theta} \kappa_{\parallel 0} \beta \frac {B_{0}}{B_m} \Bigg(\frac {P}{P_{0}}\Bigg)^{c_{1}} \left[ \frac {\Bigg(\frac {P}{P_{0}}\Bigg)^{c_{3}} + \Bigg(\frac {P_{k}}{P_{0}}\Bigg)^{c_{3}}}{ 1+ \Bigg(\frac {P_{k}}{P_{0}}\Bigg)^{c_{3}}} \right]^{\frac {c_{2 \perp\theta} - c_{1}}{c_{3}}},
\label{Eq-perpDiffTheta}
\end{equation}
where $\kappa^{0}_{\perp r}$ and $\kappa^0_{\perp\theta}$ are scaling factors that will be specified later and $f_{\perp \theta}$ is a latitude-dependent function that takes care of the anisotropy in $\kappa_{\perp \theta}$, and is given by
\begin{equation}
 f_{\perp \theta} = A^{+} \mp A^{-} \tanh \left[\frac{1}{\Delta \theta} \left({\tilde{\theta}} - 90^\circ \pm \theta_{F} \right)\right].   
\label{EqA33}
\end{equation}
The terms in this equation are defined as follows: $A^{+}$ = $({d_{\perp \theta} + 1})/{2}$, $A^{-}$ = $({d_{\perp \theta} -1})/{2}$, $\Delta \theta$ = 1/8, $\theta_{F} = 35^{\degree}$ and 
\begin{equation}
 \tilde{\theta} = \begin{cases}
 \theta & \rm\ for \rm\ \theta \geq 90^\circ, \\
 \pi - \theta & \rm\ for \rm\ \theta < 90^\circ,
 \end{cases}
 \label{EqA34}
\end{equation}
where $d_{\perp \theta}$ is a dimensionless constant that determines the enhancement factor of $\kappa_{\perp \theta}$ from its value in the equatorial plane toward the poles, with respect to $\kappa_{\parallel}$. In this study $d_{\perp \theta}$ = 6 for all particles over the entire December 2016 - January 2018 period. 
This choice means that the polar perpendicular diffusion $\kappa_{\perp\theta}$ can be enhanced toward the heliospheric poles by a factor of 6. For the motivation of this approach, see \cite{2003ApJ...594..552F, 2011AdSpR..48..300N, 2015ApJ...810..141P}.
In some cases, NLGC theory suggests a very weak, or even an independent, rigidity dependence for $\lambda_{\perp}$ below $\sim$ 10 GV \citep[e.g.][]{2010JGRA..115.3103P}, which, according to the modulation model, produces excessively large modulation above $\sim$ 1 GV. The above modelling choices are typical of the literature \citep[][]{2013LRSP...10....3P}. Later, we will discuss how some of these assumptions may need to change.

\subsubsection{Spatial Dependence of the Diffusion Coefficients  \label{radial}}
The analytical model describing the Parker spiral for radial distances $r$ $\geq$ $r_{\odot}$ \citep{1958PhRv..110.1445P}, modified here to incorporate the polarity structure of the heliospheric
current sheet \citep[see][]{1981ApJ...243.1115J,1983ApJ...265..573K}, is given by the expression

\begin{equation}
\B = B_{0}A\left(\frac{r_{0}}{r}\right)^{2}[1-2H(\theta - \theta^{'})](\e_{r}-\tan\psi\, \e_{\phi}),
\label{Eq-HMF}
\end{equation}     
where $\e_{r}$ and $\e_{\phi}$ are unit vectors in the radial and azimuthal directions, $A$ = $\pm$1, expresses the polarity phase of the Sun ($A>0$: positive polarity, $A<0$: negative polarity); $B_{0}$ is the magnitude of the HMF at the Earth (i.e. $r_{0}$ = 1 AU) and the spiral angle $\psi$ is the angle between the radial direction and a HMF field line at any given position defined by
\begin{equation}
  \tan \psi = \Omega \frac {(r - r_{\odot})} {|\vsw|} \sin\theta, 
\label{psi1}
\end{equation}  
with $\Omega$ = 2.67$\times$10$^{-6}$ rad s$^{-1}$ being the average angular rotation speed of the Sun. The Heaviside function $H$ in equation (\ref{Eq-HMF}) determines the polarity of the magnetic field which causes the HMF to change direction across the HCS,
with $\theta^{'}$ being the polar position of the HCS. The basic HMF structure resembles that of Archimedean spirals traversing cones of constant heliographic latitude. Beyond the position of the TS, the HMF lines are compressed as a result of the slower moving SW in the heliosheath region \citep[see e.g.][]{2024ApJ...961...21L}. During an $A>0$ polarity cycle, the HMF lines will be directed outward in the Northern hemisphere, and inward in the Southern hemisphere (with opposite directions for an $A < 0$ cycle). In general, the spiral angle $\psi$ in the equatorial plane at Earth is typically 45$^{\circ}$, and increases with distance to almost 90$^{\circ}$ at $r$ $\gtrsim$ 10 AU.

The magnitude of the magnetic field, from equation (\ref{Eq-HMF}), is given by

\begin{equation}
B = B_{0}\left(\frac{r_{0}}{r}\right)^{2}\sqrt{1 + (\tan \psi)^{2}.}
\label{EqA54}
\end{equation} 
from which it is evident that $B$ decreases as $r^{-2}$ at the poles. It is known, however, that the radial magnetic field at the poles is not in perfect equilibrium due to the presence of strong turbulent convection. These motions result in transverse magnetic field components in the polar regions which regularly lead to deviations from the smooth Parker field geometry \citep{1989GeoRL..16....1J,1996JGR...101..395F}. While the dynamics of changes to the radial magnetic field at the poles are complex, the effects of producing a more tightly wound spiral and larger spiral angles at larger distances can be incorporated using the adjustment proposed by \citet[][]{1991ApJ...370..435S}. They suggest that equation (\ref{psi1}) be modified to 
\begin{equation}
    \tan \psi' = \Omega \frac {(r - r_{\odot})} {|\vsw(r,\theta)|} \sin\theta -\frac{r|\vsw(r_b,\theta)|}{r_b|\vsw(r,\theta)|}\frac{B_T(r_b)}{B_R(r_b)},
\end{equation}
where $B_T/B_R$ is the ratio of the azimuthal magnetic field component to the radial component and $r_b$ is some radial distance from the center of the Sun. Here, we follow \citet[][]{2016AdSpR..57.1965R} in setting $r_b=20 r_\odot$ and $B_T(r_b)/B_R(r_b)\approx -0.02$. The HMF with this modification is denoted $\B_m$ (with magnitude $B_m$, see equation \ref{EqParallelDiffusion}). For more details about this modification and its physical justification, the reader is directed to \citet[][]{1991ApJ...370..435S} and \citet[][]{2016AdSpR..57.1965R}.

\subsection{Particle Drifts  \label{drift}}

The significance of particle drifts, not included in the original derivation of the transport equation, was not realized until \citet{1977ApJ...213..861J} pointed out that this process might contribute to CR modulation. This was later confirmed by modulation models that included drift effects \citep[][]{1983ApJ...265..573K, 1985ApJ...294..425P}.
The global background HMF induces drift motions in GCRs, which originate as a result of gradients in the HMF magnitude, the curvature of the field and any sudden changes in the field direction, like those found
at the HCS \citep[e.g.][]{1989ApJ...339..501B}, in addition to a charge asymmetry as a result of the sensitivity of drifts to the HMF polarity. The pitch angle-averaged guiding center drift velocity can be written as
\begin{equation}\label{gen_drift_vel}
    \langle\vv_d\rangle = \nabla\times\kappa_A\frac{\B}{B},
\end{equation}
where $\kappa_A$ is the global drift coefficient that is discussed below. If we write $\B/B = [1-2H(\theta-\theta')]\e_B$, where $\e_B$ is a unit vector parallel to $\B$, we can re-express equation (\ref{gen_drift_vel}) as
\begin{eqnarray}
    \langle\vv_d\rangle &=& [1-2H(\theta-\theta')]\nabla\times\kappa_A\e_B + \frac{2\kappa_A}{r}\delta(\theta-\theta')\e_B\times\e_\theta\nonumber\\
    &=& [1-2H(\theta-\theta')]\langle\vv_d\rangle_{\rm gc} + \delta(\theta-\theta')\langle\vv_d\rangle_{\rm HCS},\label{H_d}
\end{eqnarray}
where $\langle\vv_d\rangle_{\rm gc}$ represents the average drift velocity due to gradients and curvature in the HMF and $\langle\vv_d\rangle_{\rm HCS}$ represents the average drift velocity due to current sheet drift because of a switch in magnetic polarity over the HCS. The derivative of the Heaviside function leading to the delta function in equation (\ref{H_d}) is taken in the sense of distributions.

When modelling the drift velocity, we follow previous work \citep[][]{2000JGR...10527447B,2015AdSpR..56.1525N,2016AdSpR..57.1965R}  and write
\begin{equation}
\kappa_{A}= \kappa_{A0}  \frac {\beta P} {3B_{m}} \frac {(P/P_{A0})^{2}}{1+(P/P_{A0})^{2}},
\label{Eq-Drift}
\end{equation}    
where $\kappa_{A0}\in[0,1]$ is a dimensionless constant parameterizing the amount of drift reduction due to diffusive scattering and $P_{A0}$ (in GV) is a quantity added on dimensional grounds. Equation (\ref{Eq-Drift}) represents a deviation from weak scattering, with particles with rigidity below $P_{A0}$ having drifts that are progressively reduced with respect to the weak scattering case (in the weak scattering limit, $\kappa_A\rightarrow\kappa_{A0}\beta P/(3B_m)$). For more details of the properties of equation (\ref{Eq-Drift}) and its relation to observations, we direct the reader to \cite{2013LRSP...10....3P} and references within.

\subsection{vLIS}
The \emph{very local interstellar spectrum},  or vLIS, describes the spectrum of protons at the edge of the heliosphere, which we take to be 122 AU. All modulations results relate to how the proton spectrum changes relative to the vLIS. In this work, we use the same vLIS as that of \citet{2023ApJ...953..101A}, which, following \cite{2019ApJ...878...59B}, has the form
\begin{equation}
 J_p(\Gamma) = \frac{2620}{\beta^2}\Gamma^{1.1} \left(\frac{\Gamma^{0.98}+0.7^{0.98}}{1+0.7^{0.98}}\right)^{-4} + 30\Gamma^2\left(\frac{\Gamma+8}{9}\right)^{-12},   
\end{equation}
where $J_p$ is the GCR intensity (with units of particles m$^{-2}$ s$^{-1}$ sr$^{-1}$ (GeV/nuc)$^{-1}$) and $\Gamma =E/E_0$ where $E$ is the kinetic energy (GeV/nuc) and $E_0=1$ GeV/nuc. For an overview, see \citet{2021Physi...3.1190P} and references therein.

\section{Model 1: a long-term modelling approach}
The transport framework described in Section 2 has been shown to successfully reproduce GCR modulation over a complete solar cycle \citep[e.g.][]{2023ApJ...953..101A}. As a starting point, we therefore apply this long‑term modulation approach to the 2017 depression, treating the event in the same manner as other intervals within the declining phase of Solar Cycle 24. This approach provides a well‑defined baseline against which the physical consistency of the resulting transport solutions can be assessed.

We emphasize that the purpose of this section is not merely to reproduce the AMS‑02 observations which, as shown below, can be achieved to high precision, but to examine whether a conventional parameterization for long-term modulation yields transport behaviour that remains physically consistent when applied to a mid-term event like the 2017 depression. As will be demonstrated, although this approach accurately captures the observed proton intensities, it leads to an inferred rigidity dependence of particle MFPs that deviates significantly from basic turbulence theory.


We make use of the the 1-day resolution observations reported by \cite{2021PhRvL.127A1102A} for protons. In line with \citet{2023ApJ...953..101A}, we fix the parameters $\kappa_{A0}=0.9$ and $P_{A0}=0.9$ GV in equation (\ref{Eq-Drift}). The first of these relates to the drift coefficient corresponding to a 90\% weak scattering gradient and curvature drifts. For the second parameter, this describes a cut-off rigidity, below which  there is greater deviation from weak scattering. We set $\kappa^0_{\perp r}=\kappa^0_{\perp \theta}=0.02$, which is a common assumption in many previous works \citep[including][]{2023ApJ...953..101A,2021ApJS..257...48S,CortiICRC2023}.  The parameter $c_3$ is also fixed.

For each Carrington rotation considered, the code is run iteratively by varying the model parameters until the root mean square (RMS) value, across the rigidity and time ranges considered, of the percentage difference between the model and observational differential intensities is less that 5\%. As will be shown later, this approach results in RMS values lower than this threshold and the greatest percentage difference at a particuar Carrington rotation is less than 7\%.

Given the above model assumptions, and fitting the model to match differential intensity spectra data from AMS-02 (which will be displayed shortly), a full list of model parameters is displayed in Table \ref{Table1}.

\begin{deluxetable*}{lccccccccccccccc}[h]
\tablecaption{Summary of the calculated intrinsic and modulation parameters used to reproduce proton observations for the time 14 December 2016 (CR2185) to 26 January 2018 (CR2199).\label{Table1}}
\tablewidth{0pt}  
\tablehead{
  \colhead{Time} & 
  \colhead{HMF} &
  \colhead{$\alpha$} &
  \colhead{$d_{\rm TS}$} &
  \colhead{$\kappa_{\parallel 0}$} &
  \colhead{$\kappa_{A0}$} &
  \colhead{$P_{A0}$} &
  \colhead{$c_{1}$} &
  \colhead{$c_{2\parallel}$} &
  \colhead{$c_{2\perp r}$} &
  \colhead{$c_{2\perp\theta}$} &
  \colhead{$c_{3}$} &
  \colhead{$P_{k}$} &
  \colhead{$d_{\perp\theta}$} \\
  \colhead{} & 
  \colhead{(nT)} & 
  \colhead{(degrees)} &
  \colhead{(AU)} & 
  \colhead{area/time} &
  \colhead{} &
  \colhead{(GV)} &
  \colhead{} &
  \colhead{} &
  \colhead{(\% $c_{2\parallel}$)} &
  \colhead{(\% $c_{2\parallel}$)} &
  \colhead{} &
  \colhead{(GV)} &
  \colhead{} 
}
\startdata
{2185}  & 6.11  & 26.72 & 83.0 & 66.95 & 0.90 & 0.90 & 1.11 & 1.60 & 0.66 & 0.65 & 2.50 & 2.90 & 6.0    \\
{2186} & 6.00  & 25.87 & 82.50 & 68.61 & 0.90 & 0.90 & 1.05 & 1.60 & 0.66 & 0.65 & 2.50 & 3.0 & 6.0  \\
{2187} & 5.94  & 24.92 & 82.50 & 70.63 & 0.90 & 0.90 & 1.01 & 1.60 & 0.66 & 0.65 & 2.50 & 3.10 & 6.0    \\
{2188} & 5.83  & 23.53 & 82.50 & 70.77 & 0.90 & 0.90 & 0.96 & 1.60 & 0.66 & 0.65 & 2.50 & 3.20 & 6.0   \\
{2189} & 5.74  & 22.49 & 82.0 & 69.59 & 0.90 & 0.90 & 0.96 & 1.60 & 0.66 & 0.65 & 2.50 & 3.00 & 6.0  \\
{2190} & 5.70  & 21.80 & 82.0 & 72.30 & 0.90 & 0.90 & 0.96 & 1.60 & 0.66 & 0.65 & 2.50 & 3.10 & 6.0   \\
{2191} & 5.64  & 21.19 & 82.0 & 72.44 & 0.90 & 0.90 & 0.94 & 1.58 & 0.66 & 0.65 & 2.50 & 3.30 & 6.0   \\
{2192} & 5.56  & 20.56 & 82.0 & 70.77 & 0.90 & 0.90 & 0.92 & 1.58 & 0.65 & 0.65 & 2.50 & 3.30 & 6.0  \\
{2193} & 5.31  & 19.48 & 81.50 & 63.89 & 0.90 & 0.90 & 0.92 & 1.58 & 0.65 & 0.65 & 2.50 & 3.0 & 6.0  \\
{2194} & 5.30  & 18.99 & 81.50 & 62.91 & 0.90 & 0.90 & 0.92 & 1.58 & 0.65 & 0.65 & 2.50 & 3.0 & 6.0  \\
{2195} & 5.27  & 18.66 & 81.50 & 60.13 & 0.90 & 0.90 & 0.90 & 1.58 & 0.66 & 0.65 & 2.50 & 2.80 & 6.0  \\ 
{2196} & 5.29  & 18.05 & 81.50 & 65.00 & 0.90 & 0.90 & 0.90 & 1.60 & 0.665 & 0.65 & 2.50 & 2.70 & 6.0 \\
{2197} & 5.29  & 17.34 & 81.50 & 70.56 & 0.90 & 0.90 & 0.89 & 1.60 & 0.665 & 0.65 & 2.50 & 2.80 & 6.0 \\
{2198} & 5.28  & 16.29 & 81.0 & 73.00 & 0.90 & 0.90 & 0.88 & 1.60 & 0.665 & 0.65 & 2.50 & 2.80 & 6.0\\
{2199} & 5.24  & 15.31 & 81.0 & 74.79 & 0.90 & 0.90 & 0.88 & 1.60 & 0.665 & 0.65 & 2.50 & 2.80 & 6.0  \\
\enddata
\tablecomments {Time is displayed by Carrington rotation number, 2185 (14 December 2016 - 09 January 2017) - 2199 (30 December 2017 - 27 January 2018). The distance to the termination shock is denoted $d_{\rm TS}$. All other parameters are defined in the main text.}
\end{deluxetable*}
The parameter $c_1$ decreases with time, which is indicative of long-term modulation. The remaining parameters in Table \ref{Table1} vary with time in a non-monotonic fashion, which is necessary in order to reproduce the observed behaviour of the 2017 GCR depression. Whilst some parameters, like $\kappa_{\parallel 0}$, exhibit minima that correspond to the minimum of the depression, it is important to take into account nonlinear relations with other parameters, such as $P_k$ and the $c$-powers for diffusion, in order to develop a complete understanding of the key underlying physics. For the period under study, the Sun was in a positive HMF polarity phase ($A>0)$. Thus, for protons, we set $A=1$. 

\subsection{Differential intensity and flux variation}
\begin{figure}
    \centering
   \includegraphics[width=0.70\linewidth]{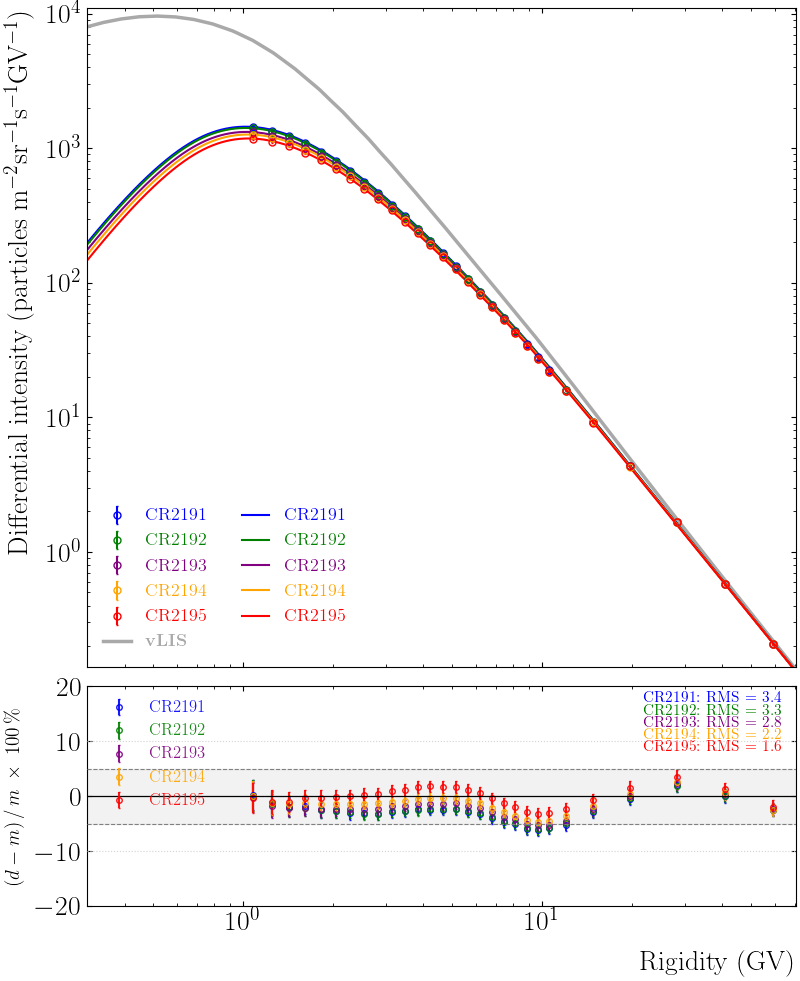}
   \caption{Differential Intensity of galactic protons at Earth (1 AU) during the Carrington rotations 2191 - 2195. The lines are model results and circles are AMS-02 observations adopted from \cite{2021PhRvL.127A1102A}. The grey line shows the shape of vLIS. The accuracy of the model fit is shown below where the relative percentage difference between the data ($d$) and the model ($m$). The root mean square (RMS) errors per Carrington rotation are also highlighted.}  
    \label{Figure1:DI}
\end{figure}

Based on the above parameterization, Figure \ref{Figure1:DI} shows the modelled differential intensity spectra of GCR protons at the Earth for Carrington rotations 2191–2195, together with AMS‑02 observations. The corresponding vLIS is shown for reference. The parameter values listed in Table \ref{Table1} are chosen to reproduce the observed spectral shapes and differential intensities across the full range of measured rigidities.


At lower rigidities, where observational data are not available, the shape of the differential intensity spectra at the Earth is influenced by the fact that protons are highly susceptible to adiabatic cooling, which is the reason for the steady slopes of these modulated spectra (see also \cite{2020Ap&SS.365..182N}). For observational evidence of this behaviour, see, for example,  \cite{2018ApJ...854L...2M}.

\begin{figure}[!ht]
    \centering
    \includegraphics[width=0.9\linewidth]{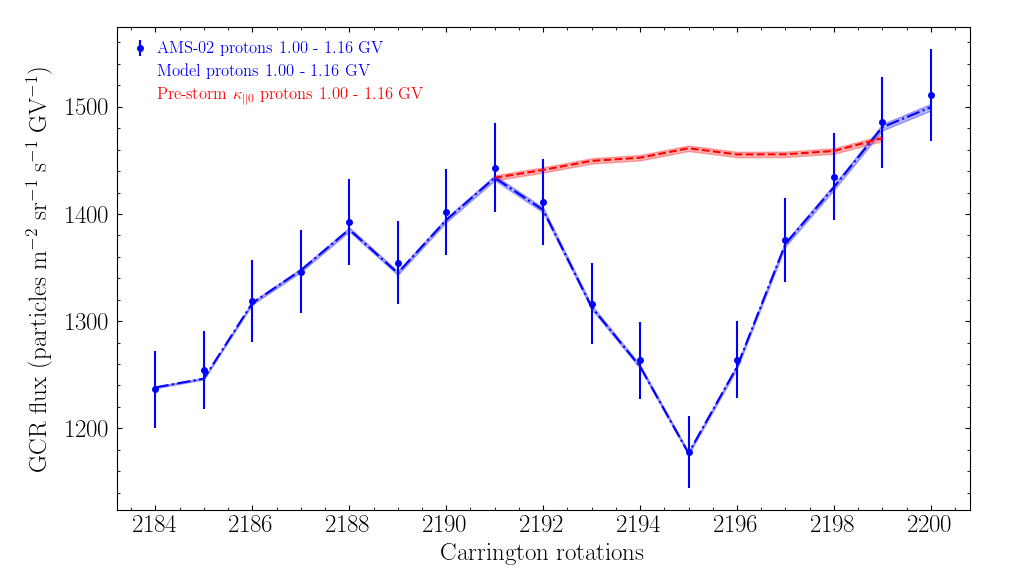}
    \caption{Time variation of modeled galactic proton flux for the rigidity range 1.0 - 1.16 GV along with standard error of mean (SEM = standard deviation $\sigma/\sqrt{n}$) as shading around solid line, here compared with corresponding AMS-02 observations of the same rigidity range for the time period CR2184 - CR2200. The meaning of the red dashed line is discussed in the main text.}  
     \label{Fig2:TimeVariation}
\end{figure}
The temporal evolution of the proton flux for the lowest rigidity bin measured by AMS‑02 is shown in Figure \ref{Fig2:TimeVariation}. The model accurately reproduces both the timing and magnitude of the depression, as well as the subsequent recovery toward near‑minimum conditions. This agreement highlights the ability of the long‑term modulation framework to capture the integrated impact of the transient CME activity on the observed proton fluxes. In this sense, the model performs at least as well as previous steady‑state and time‑dependent approaches applied to this interval.


In the model, we can also experiment with how changing a particular part of the physics affects the modulation profile. For example, if we fix $\kappa_{\parallel 0}$ = 72.44, i.e. the value just before the depression, and maintain this value from CR2191 to CR2199, whilst keeping all other values in Table \ref{Table1} as before, the result is indicated by the red line in Figure \ref{Fig2:TimeVariation}. Essentially, the proton flux continues its recovery toward solar minimum modulation as if there were no CMEs disturbing it. This is a test of the model's consistency as this behaviour is what would be expected if there were no CMEs - there would be no disturbance due to the CMEs and thus no depression. This result further highlights the importance of varying a quantity like $\kappa_{\parallel0}$ in time, in addition to changing the HMF magnitude and the HCS tilt angle.

However, it is important to emphasize that the quality of the spectral and temporal fits alone does not uniquely constrain the underlying transport physics. The modulation profile primarily reflects the net effectiveness of competing transport processes acting over the heliosphere, allowing correlated adjustments of diffusion coefficients and rigidity dependences to compensate for one another. As a result, transport parameterizations can reproduce the observed intensities even when the implied rigidity dependence of particle scattering deviates from standard physical expectations. To assess whether the long‑term parameterization (in Table \ref{Table1}) yields physically consistent transport behaviour during this transient event, it is therefore necessary to examine the inferred rigidity dependence of the diffusion and drift scales in more detail.

\subsection{MFPs and drift scales}
\begin{figure}[!ht]
    \centering
    \includegraphics[width=1.0\linewidth]{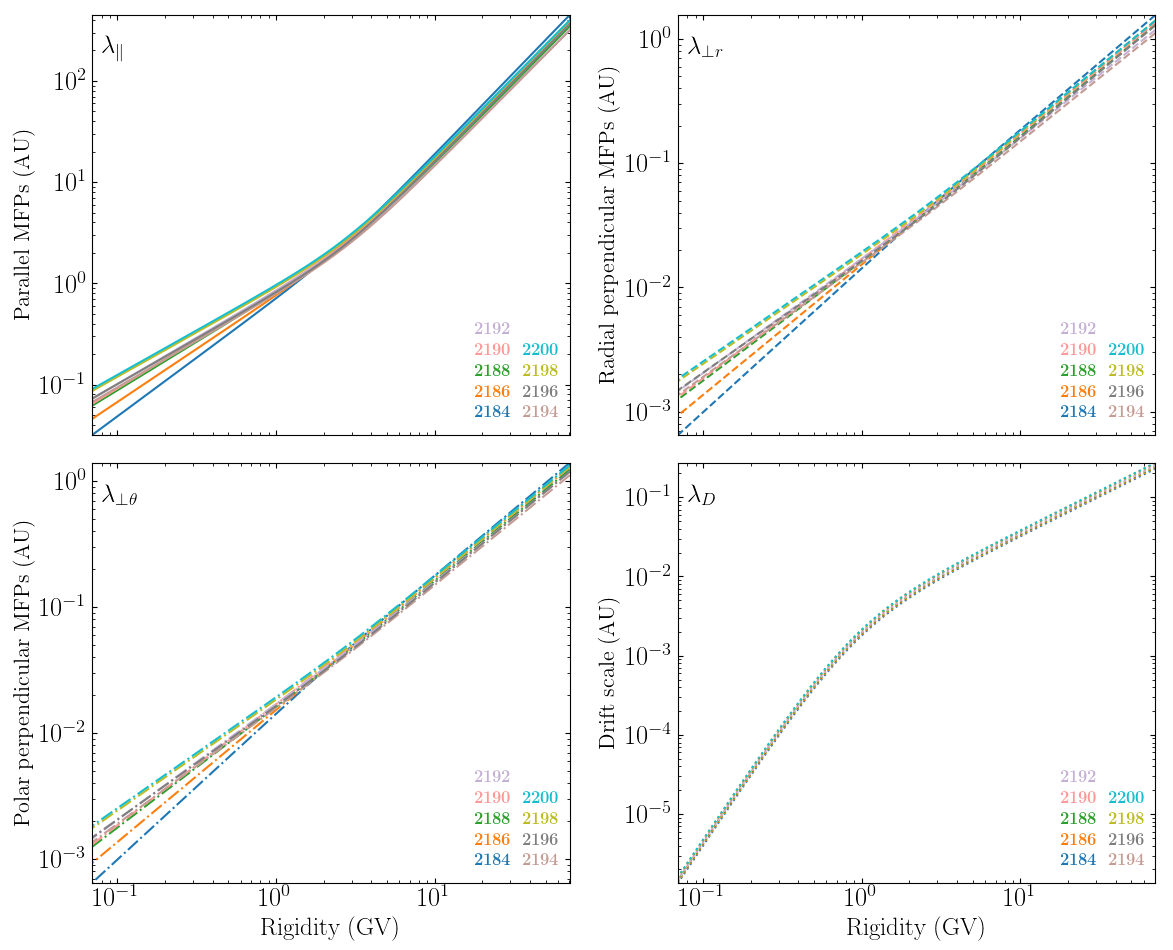}
    \caption{Rigidity dependence of the parallel ($\lambda_{\parallel}$) and two perpendicular ($\lambda_{\perp r}$, $\lambda_{\perp\theta}$) MFPs, and the drift scale ($\lambda_D$) in units of AU for protons for alternate CRs from 2184 - 2200. 
    }  
     \label{Fig3:MFPs}
\end{figure}

As discussed above, the rigidity dependence of diffusion MFPs and drift scales plays a central role in determining the modulation of GCRs. To examine the physical implications of the long‑term modulation parameterization, Figure \ref{Fig3:MFPs} shows the rigidity dependence of the parallel MFP $\lambda_\|$, the perpendicular MFPs $\lambda_{\perp r}$ and $\lambda_{\perp\theta}$ and the drift scale $\lambda_D$ at Earth for selected Carrington rotations spanning the 2017 depression. The functional forms of the quantities displayed in Figure \ref{Fig3:MFPs}, as well as their numerical values, are similar to those of \citet{2023ApJ...953..101A}.


\begin{figure}[!ht]
   \centering
    \includegraphics[width=0.7\linewidth]{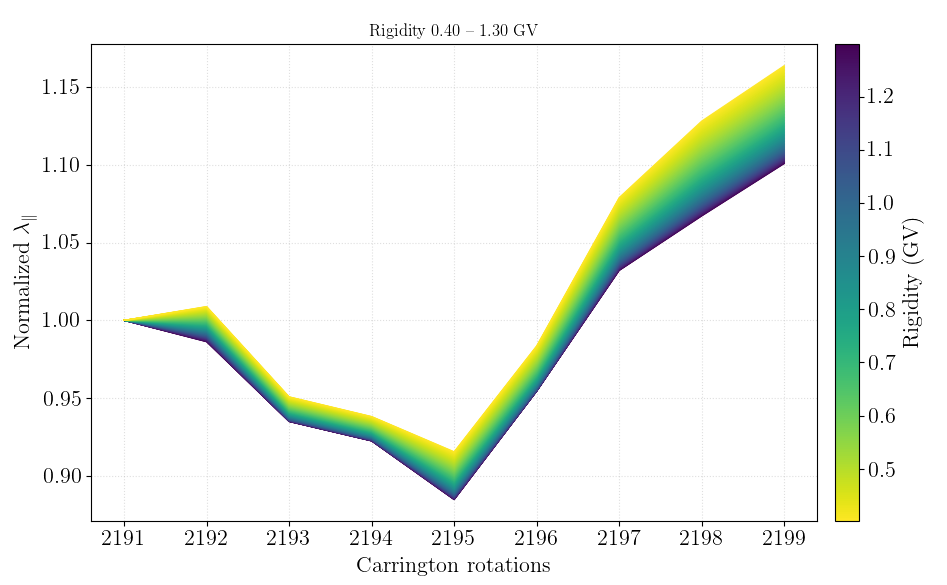}
        \vspace{0.1cm} 
    \includegraphics[width=0.7\linewidth]{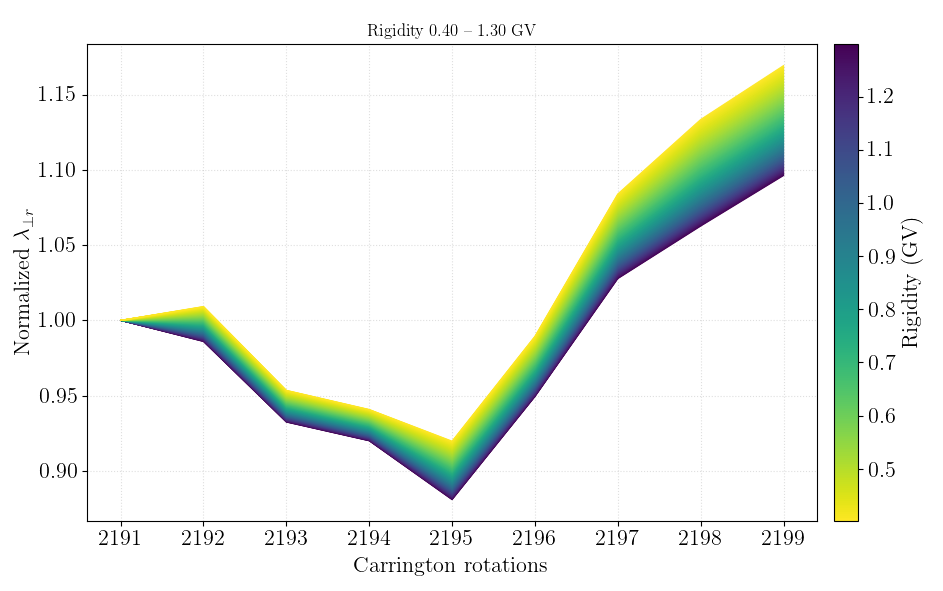}
        \caption{Time dependence in parallel ($\lambda_{\parallel}$; top)  and perpendicular ($\lambda_{\perp r}$; bottom) MFPs of protons for rigidities 0.40 - 1.30 GV, over the 2017 depression period CR2191 - CR2198, normalized with respect to CR2191.}  
     \label{Fig4:3D-MFPs}
\end{figure}

The corresponding temporal evolution of the parallel and (radial) perpendicular MFPs for a range of rigidities is shown in Figure \ref{Fig4:3D-MFPs}.

During the development of the depression, Figure \ref{Fig4:3D-MFPs} reveals that the relative reduction of MFPs is stronger for higher‑rigidity protons than for lower‑rigidity protons. In particular, the parallel and radial perpendicular MFPs at higher rigidities decrease more sharply through the onset and peak of the depression. This behaviour implies that higher‑rigidity particles experience comparatively stronger modulation than lower‑rigidity particles during this interval. Such a rigidity ordering corresponds to an effective scattering regime that is opposite to that expected for heliospheric turbulence dominated by large‑scale fluctuations.

Under conventional assumptions, higher‑rigidity particles are expected to undergo weaker resonant scattering and therefore possess larger mean free paths than lower‑rigidity particles. The behaviour inferred from this model, therefore, challenges standard expectations and is described, in an inferred sense, as corresponding to an inverted rigidity dependence of particle scattering. We stress that this inversion is not a direct measurement of the heliospheric magnetic‑field power spectrum, but rather a diagnostic implication arising from the fitted rigidity dependence of the transport coefficients within the adopted modulation framework.


Importantly, this inferred inversion occurs despite the excellent agreement between modelled and observed proton intensities demonstrated in Figures \ref{Figure1:DI} and \ref{Fig2:TimeVariation}. This apparent contradiction can be understood in terms of dominant balance in the steady‑state Parker transport equation. When averaged over a Carrington rotation, the transport solution is governed by a balance between outward solar‑wind convection and a combination of diffusion, drifts and adiabatic energy changes, as indicated in equation (\ref{balance}), 
\begin{equation}\label{balance}
    0\approx -\vsw\cdot\nabla f   +\nabla\cdot[\kappa_\|\nabla_\| f + \kappa_\perp\nabla_\perp f] - \langle\vv_D\rangle\cdot\nabla f + \frac {1}{3} (\nabla \cdot \vsw) \frac {\partial f} {\partial  \ln  p}.
\end{equation}
As a consequence, different combinations of $\kappa_{\parallel}$, $\kappa_{\perp}$ and their associated rigidity slopes (see Section 2 for more details) can satisfy the same dominant balance at a given rigidity, allowing for a physically inconsistent rigidity ordering of MFPs to persist without degrading the agreement with observed fluxes. Thus, agreement with observed intensities alone is insufficient to guarantee physically consistent rigidity‑dependent transport during transient modulation events, and so motivates the need for additional physical constraints beyond goodness‑of‑fit considerations.

The appearance of this behaviour in a long‑term modulation parameterization, together with similar results reported in both steady‑state and time‑dependent studies \citep[][]{CortiICRC2023,2021ApJS..257...48S}, is suggestive of a limitation of commonly adopted model assumptions when applied to CME‑dominated hybrid intervals, like the 2017 depression, rather than evidence for a genuine reversal of heliospheric turbulence properties. In the following section, we, therefore, explore whether relaxing one such assumption, motivated by the physical conditions of the 2017 depression, can constrain the transport solutions and restore the expected rigidity ordering of the MFPs.



\section{Model 2: adapting for ``hybrid'' conditions}

 The results of Section 3 demonstrate that treating the 2017 depression as a purely long‑term modulation interval leads to transport solutions that reproduce the observed proton intensities but imply a physically inconsistent rigidity dependence of particle scattering. As argued above, this behaviour is a consequence of the nonlinear nature of the Parker transport equation. Different combinations of diffusion and drift coefficients can produce similar modulation profiles because the observed flux reflects the net balance of several competing transport processes. Consequently, agreement with observations does not by itself guarantee that the inferred rigidity dependence of the transport parameters is physically realistic.

The purpose of this section, therefore, is not to improve the quality of the fit to the AMS-02 data, but to investigate whether an additional physically motivated constraint can reduce the degeneracy of the transport solution space and eliminate the inferred rigidity-ordering inconsistency identified in Section 3. As demonstrated by  \citet{2025ApJ...981..174A}, the interval CR2191--CR2197 was characterized by repeated CME-driven disturbances that continually interrupted the recovery of the proton flux. Such disturbances are commonly interpreted as diffusion barriers or magnetic obstacles that inhibit cosmic-ray penetration into disturbed heliospheric regions \citep[][]{Lockwood1971,Wibberenz1998,Cane2000,Dumbovic2020,Dumbovic2022}. Since the transport coefficients used here represent effective Carrington-averaged quantities rather than instantaneous local turbulence parameters, the repeated occupation of successive Carrington rotations by strong CME-driven disturbances implies an effective transport regime that differs from that of the surrounding near-minimum heliosphere. We, therefore, investigate whether a modest reduction in the effective ratio \(\kappa_{\perp r}/\kappa_\parallel\), motivated by the diffusion-barrier interpretation of CME-driven cosmic-ray depressions, can provide an additional physical constraint on the transport solution space and thereby eliminate the inferred rigidity-ordering inconsistency identified in Section~3.

Performing this change, we find that modifying the polar component of the perpendicular diffusion coefficient, $\kappa_{\perp\theta}$, has a negligible impact on the modulation results for the 2017 depression, whereas changes to the radial component, $\kappa_{\perp r}$, produce a substantial effect. This difference arises because the observed proton fluxes at the Earth are primarily sensitive to radial transport through the heliosphere, particularly during intervals dominated by outward‑propagating CMEs. While enhanced polar perpendicular diffusion plays an important role in reproducing the weak latitudinal gradients observed by Ulysses and related missions, it contributes only indirectly to the radial modulation experienced by particles reaching 1 AU. In contrast, $\kappa_{\perp r}$ directly controls the efficiency with which particles can diffuse across the Parker spiral field while propagating radially inward against solar‑wind convection. Consequently, reducing $\kappa_{\perp r}$ during the CME‑dominated phase significantly alters the relative modulation of particles at different rigidities, suppressing compensating cross‑field transport and restoring the expected rigidity ordering of the MFPs. We, therefore, leave $\kappa_{\perp\theta}$ at its long‑term value and focus exclusively on changing $\kappa_{\perp r}$ as a minimal and physically motivated adjustment to the standard parameterization.

\begin{figure}[!ht]
    \centering
    \includegraphics[width=0.6\linewidth]{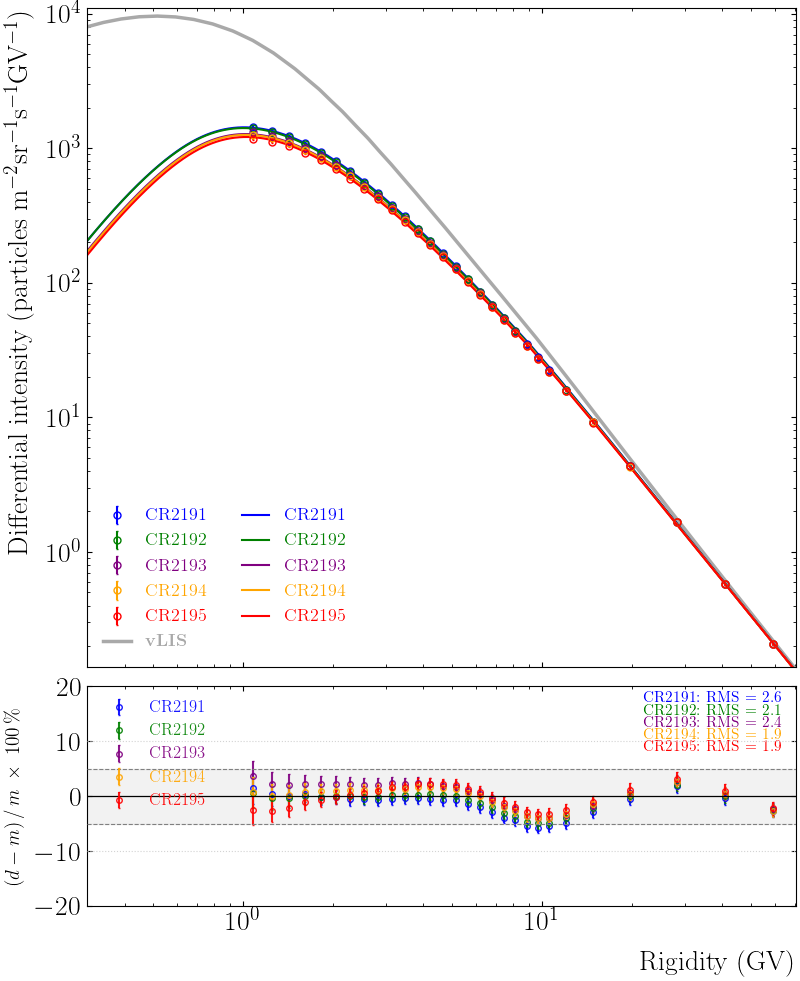}
    \caption{Model 2 results (solid lines) of the differential intensity of galactic protons at the Earth (1 AU) during Carrington rotations 2191 - 2195, using the new parameters for the 2017 GCR depression displayed in Table \ref{Table2}. AMS-02 observations \citep{2021PhRvL.127A1102A} are shown as circles and the vLIS is displayed as a grey line. The accuracy of the model fit is shown below where the relative percentage difference between the data ($d$) and the model ($m$). The root mean square (RMS) errors per Carrington rotation are also highlighted.}  
     \label{Fig5:Newspectra}
 \end{figure}

\begin{deluxetable*}{lcccccccccccccccc}[h]
\tablecaption{Summary of the calculated intrinsic and modulation parameters (changed parameters are in bold) used to reproduce proton observations for the CR2190 - CR2198.\label{Table2}}
\tablewidth{0pt}  
\tablehead{
  \colhead{Time} & 
  \colhead{HMF} &
  \colhead{$\alpha$} &
  \colhead{$d_{\rm TS}$} &
  \colhead{$\kappa_{\parallel 0}$} &
  \colhead{$\kappa^{0}_{\perp r}$} &
  \colhead{$\kappa_{A0}$} &
  \colhead{$P_{A0}$} &
  \colhead{$c_{1}$} &
  \colhead{$c_{2\parallel}$} &
  \colhead{$c_{2\perp r}$} &
  \colhead{$c_{2\perp\theta}$} &
  \colhead{$c_{3}$} &
  \colhead{$P_{k}$} &
  \colhead{$d_{\perp\theta}$} \\
  \colhead{} & 
  \colhead{(nT)} & 
  \colhead{(degrees)} &
  \colhead{(AU)} & 
  \colhead{area/time} &
 \colhead{} &
  \colhead{(GV)} &
  \colhead{} &
  \colhead{} &
  \colhead{} &
  \colhead{(\% $c_{2\parallel}$)} &
 \colhead{(\% $c_{2\parallel}$)} &
  \colhead{} &
  \colhead{(GV)} &
  \colhead{} 
}
\startdata
{2190} & 5.70  & 21.80 & 82.0 & 72.30 & 0.02& 0.90 & 0.90 & 0.96 & 1.60 & 0.66 & 0.65 & 2.50 & 3.10 & 6.0   \\
{2191} & 5.64  & 21.19 & 82.0 & \textbf{89.46} & \textbf{0.016}& 0.90 & 0.90 & \textbf{0.94} & 1.58 & \textbf{0.55} & 0.65 & 2.50 & 3.30 & 6.0   \\
{2192} & 5.56  & 20.56 & 82.0 & \textbf{85.88} & \textbf{0.016}& 0.90 & 0.90 & \textbf{0.93} & 1.58 & \textbf{0.55} & 0.65 & 2.50 & 3.30 & 6.0    \\
{2193} & 5.31  & 19.48 & 81.50 & \textbf{78.23} & \textbf{0.016}& 0.90 & 0.90 & \textbf{0.92} & 1.58 & \textbf{0.55} & 0.65 & 2.50 & \textbf{3.30} & 6.0  \\
{2194}  & 5.30  & 18.99 & 81.50 & \textbf{76.78} & \textbf{0.016}& 0.90 & 0.90 & \textbf{0.955} & 1.58 & \textbf{0.55} & 0.65 & 2.50 & \textbf{3.30} & 6.0   \\
{2195} & 5.27  & 18.66 & 81.50 & \textbf{71.75} & \textbf{0.016}& 0.90 & 0.90 & \textbf{0.985} & 1.58 & \textbf{0.55} & 0.65 & 2.50 & \textbf{3.30} & 6.0   \\ 
{2196} & 5.29  & 18.05 & 81.50 & \textbf{78.60} & \textbf{0.016}& 0.90 & 0.90 & \textbf{0.955} & 1.58 & \textbf{0.55} & 0.65 & 2.50 & \textbf{3.30} & 6.0  \\
{2197} & 5.29  & 17.34 & 81.50 & \textbf{85.04} & \textbf{0.016}& 0.90 & 0.90 & \textbf{0.960} & 1.58 & \textbf{0.55} & 0.65 & 2.50 & \textbf{3.30} & 6.0   \\
{2198} & 5.28  & 16.29 & 81.0 & \textbf{87.76} & \textbf{0.016}& 0.90 & 0.90 & \textbf{0.965}& 1.58 & \textbf{0.55} & 0.65 & 2.50 & \textbf{3.30} & 6.0  \\
{2199} & 5.24  & 15.31 & 81.0 & 74.79 & 0.02 & 0.90 & 0.90 & 0.88 & 1.60 & 0.665 & 0.65 & 2.50 & 2.80 & 6.0  \\
\enddata
\end{deluxetable*}
Table \ref{Table2} summarizes the revised parameterization adopted during the CME‑dominated interval CR2190–CR2198, with the parameters that differ from the long‑term model highlighted in bold. Relative to Table \ref{Table1}, the key modification is a reduction in the normalization of the radial perpendicular diffusion coefficient, $\kappa_{\perp r}$, during the depression. The value $\kappa_{\perp r}=0.016$ is the first value, steadily decreasing in increments of 0.001 from the Model 1 value of 0.02, for which there is a qualitative change in the behaviours of the MFPs relative to Model 1.  All other parameters are adjusted only as required to preserve the quality of the fit to the AMS‑02 proton observations which, as is clear from Figure \ref{Fig5:Newspectra}, is comparable to that of Model 1.

An important consequence of reducing $\kappa_{\perp r}$ is that the parallel diffusion normalization $\kappa_{\|0}$ increases during the depression relative to the long‑term parameterization, while retaining its minimum near the peak of the event at CR2195. This compensating change reflects the nonlinear coupling between diffusion coefficients required to maintain the observed modulation depth when cross‑field transport is suppressed. At the same time, the temporal evolution of the rigidity slope parameter $c_1$ is no longer strictly monotonic. This behaviour indicates that the 2017 depression is better characterized as a mid‑term modulation event, superimposed on the long‑term solar‑cycle trend, rather than as part of smooth secular evolution.

Despite these adjustments, the revised parameterization differs from the long‑term model in a limited and controlled manner.  This demonstrates that restoring physically consistent rigidity‑dependent transport does not require a wholesale restructuring of the model, but can be achieved by relaxing a single simplifying assumption in a physically motivated way.


\begin{figure}[!ht]
   \centering
   \includegraphics[width=0.9\linewidth]{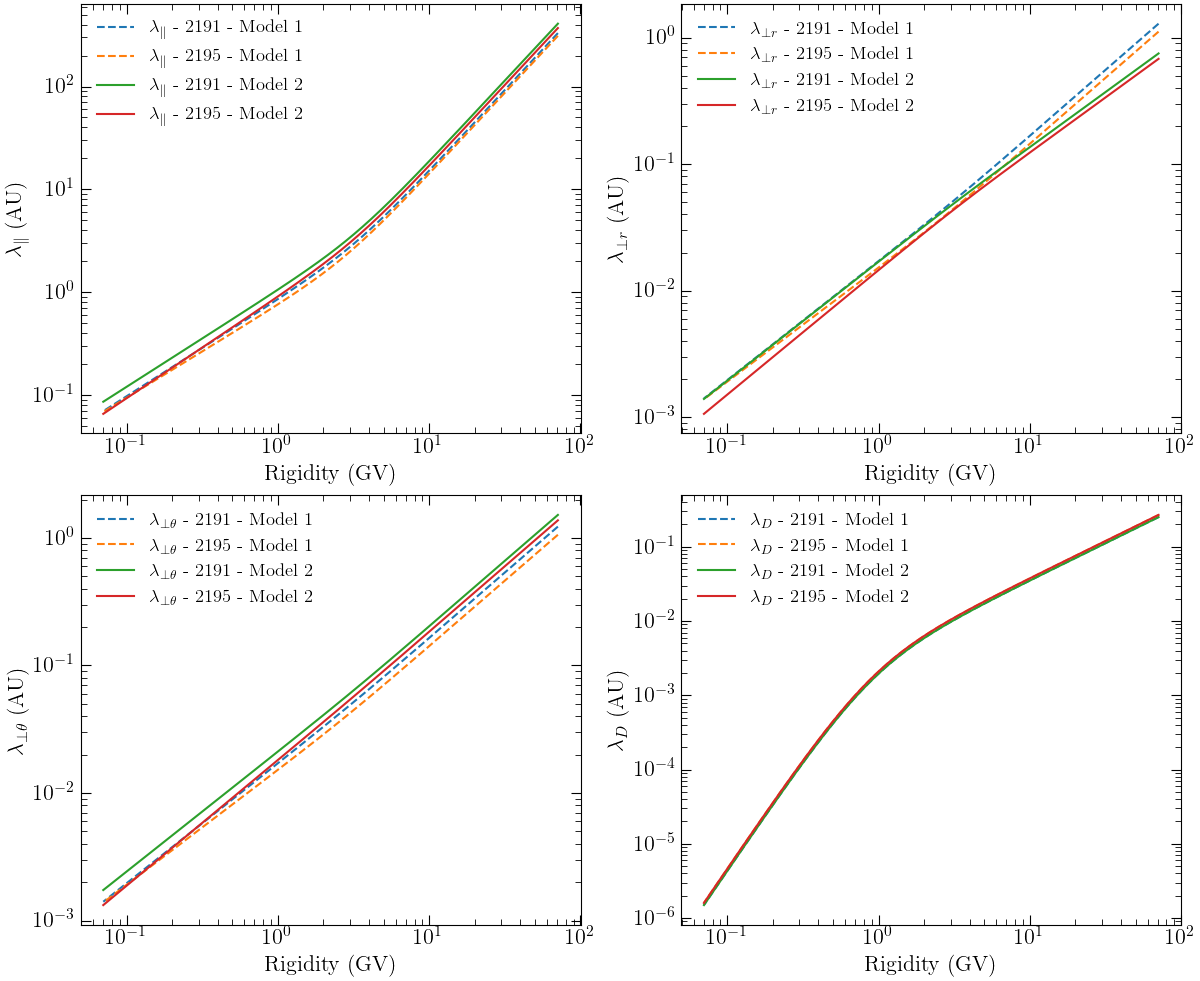}
   \caption{Rigidity dependence of MFPs and the drift scale - a comparison between Model 1 (dashed lines) and Model 2 (solid lines) for CR2191 and CR2195.}  
     \label{Fig6:MFPs-update}
\end{figure}

Figure \ref{Fig6:MFPs-update} compares the rigidity dependence of the parallel MFPs, the perpendicular MFPs  and the drift scale for Models 1 and 2 at CR2191 and CR2195, representing the onset and peak of the 2017 depression respectively. While the rigidity dependences of $\lambda_\|$, $\lambda_{\perp\theta}$, and $\lambda_D$ are broadly similar between the two parameterizations, a clearer qualitative difference is evident in the behaviour of the radial perpendicular mean free path $\lambda_{\perp r}$. In Model 2, $\lambda_{\perp r}$ exhibits a noticeably flatter rigidity dependence at higher rigidities, particularly at the peak of the depression, in closer agreement with expectations from basic theory \citep[e.g.][]{2022SSRv..218...33E}. This contrasts with Model 1, in which $\lambda_{\perp r}$ increases more steeply with increasing rigidity, contributing to the inferred inversion of the scattering regime.


\begin{figure}[!ht]
    \centering
    \includegraphics[width=0.7\linewidth]{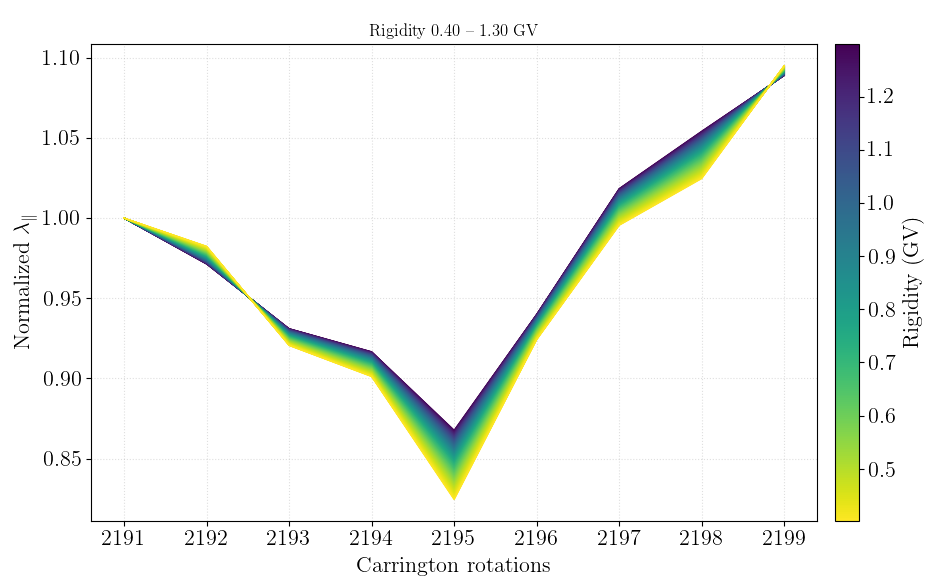}
        \vspace{0.1cm} 
    \includegraphics[width=0.7\linewidth]{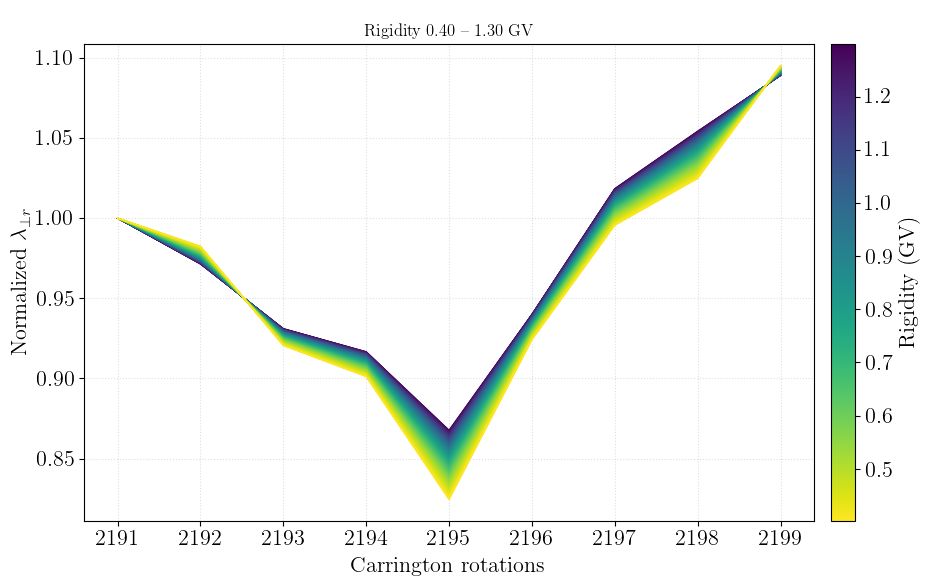}
        \caption{Time dependence in parallel ($\lambda_{\parallel}$; top)  and perpendicular ($\lambda_{\perp r}$; bottom) MFPs of protons for rigidities 0.40 - 1.30 GV, over the 2017 depression period CR2191 - CR2199, normalized with respect to CR2191.
        These values are based on the parameterization of Model 2.}  
     \label{Fig7:3D-Upd-MFPs}
\end{figure}

The impact of this difference becomes most apparent when examining the temporal evolution of the mean free paths. Figure \ref{Fig7:3D-Upd-MFPs} shows the time dependence of $\lambda_\|$ and $\lambda_{\perp r}$ for a range of rigidities, normalized to their values at CR2191. In Model 2, lower‑rigidity particles experience a more pronounced relative reduction in MFPs during the development of the depression, consistent with standard modulation theory. Moreover, as the depression subsides toward CR2198, the MFPs of lower‑rigidity particles recover more rapidly than those at higher rigidities, restoring the expected rigidity ordering throughout the recovery phase.

By contrast, in Model 1 (Figure \ref{Fig4:3D-MFPs}), higher‑rigidity particles exhibit stronger relative suppression through much of the depression, leading to the inferred inverted rigidity dependence discussed in Section 3.2. The comparison between Figures \ref{Fig4:3D-MFPs} and \ref{Fig7:3D-Upd-MFPs}, therefore, demonstrates that allowing the perpendicular‑to‑parallel diffusion ratio to vary during the CME‑dominated interval suppresses compensating cross‑field transport and constrains the dominant balance of the transport equation in a physically consistent manner. Importantly, this improvement is achieved without degrading the agreement with the observed proton intensities, indicating that the revised parameterization restores physical consistency rather than merely reshuffling the fit.


\section{Summary and conclusions}
In this work, we model the modulation of GCR proton flux during the second half of 2017, when a particularly large depression in the GCR proton flux was detected with AMS-02. We do this by applying a well-tested model \citep[e.g.,][]{2014SoPh..289..391P,2022AdSpR..69.2330N}, matched to the latest AMS-02 observations of the period. We first follow the approach of \citet{2023ApJ...953..101A} in modelling the depression as a long-term modulation event. Although we achieve a good  match between the model and the observed spectra, the results lead to a physical inconsistency in which the mean free paths (MFPs) of protons are more strongly affected for particles of higher rigidity compared to those of low rigidity. This result is consistent with an inferred rigidity dependence of particle scattering that departs from standard expectations \citep[e.g.][]{CortiICRC2023}. To address this issue, we adapt the standard model assumption of a constant ratio of perpendicular to parallel diffusion tensor coefficients, $\kappa_\perp/\kappa_\parallel$, and consider a time-dependent ratio. In particular, we reduce the value of $\kappa_\perp/\kappa_\parallel$ during the depression in order to better model this ``hybrid'' period of CME activity during otherwise near-minimum conditions. The ratio during the depression is reduced by approximately 20\% relative to its long-term value, motivated by the diffusion-barrier interpretation of CME-driven cosmic-ray depressions, while preserving agreement with the AMS-02 observations. Performing a change like this does not necessarily guarantee a physically consistent solution but it does represent a physically motivated approach to improving the model.

The effect of this small change is substantial on the model results. The new parameterization leads to a physically consistent solution, i.e. the MFPs recover the physically expected rigidity ordering from cosmic‑ray transport theory, consistent with standard turbulence‑based interpretations of heliospheric diffusion. Also, the behaviour of the radial part of the perpendicular MFP exhibits a flatter profile at higher rigidities - another property that brings the solution closer in line with basic theory \citep[e.g.][]{2022SSRv..218...33E}.

The results found here suggest that modelling $\kappa_\perp/\kappa_\parallel$ as varying in time may help to improve the results of other studies \citep[e.g.][]{2021ApJS..257...48S,CortiICRC2023} and remove some physical inconsistencies (assuming there is a physical justification for changing the ratio in time, as there is for the 2017 GCR depression). We stress that this modification does not guarantee physical consistency in all circumstances, but offers a targeted and observationally supported adjustment for a specific class of modulation events. Its applicability to other periods and to fully time‑dependent models \citep[e.g.][]{2021ApJS..257...48S} remains an important topic for future investigation.

\begin{acknowledgments}
O.P.M.A. and D.M. acknowledge support from a Leverhulme Trust grant (RPG-2023-182). D.M. also acknowledges support from a  Science and Technologies Facilities Council (STFC) grant (ST/Y001672/1) and a Personal Fellowship from the Royal Society of Edinburgh (ID: 4282). Results were obtained using the ARCHIE-WeSt High Performance Computer (www.archie-west.ac.uk) based at the University of Strathclyde. 

\end{acknowledgments}


\bibliography{sample701}{}
\bibliographystyle{aasjournalv7}



\end{document}